# A highly scalable repository of waveform and vital signs data from bedside monitoring devices

Sanjay Malunjkar, Susan Weber, Somalee Datta

Research IT, Technology & Digital Solutions, Stanford Medicine

## Abstract

The advent of cost effective cloud computing over the past decade and ever-growing accumulation of high-fidelity clinical data in a modern hospital setting is leading to new opportunities for translational medicine. Machine learning is driving the appetite of the research community for various types of signal data such as patient vitals. Health care systems, however, are ill suited for massive processing of large volumes of data. In addition, due to the sheer magnitude of the data being collected, it is not feasible to retain all of the data in health care systems in perpetuity. This gold mine of information gets purged periodically thereby losing invaluable future research opportunities. We have developed a highly scalable solution that:
1. Siphons off patient vital data on a nightly basis from on-premises bio-medical systems to a cloud storage location as a permanent archive.
2. Reconstructs the database in the cloud.
3. Generates waveforms, alarms and numeric data in a research-ready format.
4. Uploads the processed data to a storage location in the cloud ready for research.

The data is de-identified and catalogued such that it can be joined with Electronic Medical Records (EMR) and other ancillary data types such as electroencephalogram (EEG), radiology, video monitoring etc. This technique eliminates the research burden from health care systems.

This highly scalable solution is used to process high density patient monitoring data aggregated by the Philips Patient Information Center iX (PIC iX) hospital surveillance system for archival storage in the Philips Data Warehouse Connect enterprise-level database. The solution is part of a broader platform that supports a secure high performance clinical data science platform.

## Introduction

In accordance with the Stanford School of Medicine's Precision Health initiative, we are building a new data science platform for improved healthcare research. This platform comprises a data lake (STAnford medicine Research data Repository or STARR), a collection of linked research ready data warehouses from disparate clinical ancillary systems and a secure data science facility [Datta2020]. Here, we present a scalable solution that generates research ready data in the form of waveforms and vitals, from Philips patient monitoring devices.

The resulting research ready dataset STARR-Wave contains data starting from early 2017 and can be linked to Electronic Health Records (EHR) and other data such as imaging data in the STARR data lake. The solution was developed using bedside monitoring data collected from approximately 500 beds from Stanford Children's Hospital aggregated by the Philips Patient Information Center iX (PIC iX). The Philips Data Warehouse Connect (Philips DWC) application, part of the PIC iX system, stores data from Philips patient monitors, telemetry devices, and third-party devices connected to the Philips IntelliBridge family of the products in an enterprise level SQL database. Philips DWC captures patient vital numerics such as Heart Rate (HR), Blood Pressure (BP), Pulse Oximetry ($SpO_2$); alarms and alerts; and continuous waveforms such as Electrocardiogram (ECG) and invasive pressures. It contains data collected from clinical units such as Post-Anesthesia Care Unit (PACU), Intensive Care Unit (ICU), Emergency Department (ED), Magnetic Resonance Imaging (MRI), Neonatal Intensive Care Unit (NICU), Operating Room (OR), etc. Currently STARR-Wave only includes data from our 500 bed pediatric hospital; our goal is to include our adult hospital data in the future.

Researchers at Stanford have successfully used the Stanford vitals data in studies ranging from method development [Miller2018], quantitative analysis [Scala2020] to machine learning [Hannun2019]. The "Resuscitations Outcomes" committee routinely uses archived data for retrospective review of in-hospital cardiac/pulmonary arrests and resuscitations. Researchers elsewhere have also successfully used vitals data in quality improvement [Walsh2016], clinical trials [Mitra2020], method development [Charlton2016], quantitative analysis [Joshi2016, Pelter2020] and machine learning [Ansari2016]. Our STARR-Wave solution supports a growing number of new studies at Stanford. In clinical trials, one research team is testing a wearable vital sign device against vitals data from ED, and another team is executing a double blind randomized controlled trial for dose efficacy of dexmedetomidine used with propofol for procedural sedation of pediatric subjects undergoing MRI scans. A machine learning team is focusing on alarm fatigue reduction using alarms each month for validation. Another study is focusing on understanding the frequency and length of pauses in chest compression that occur in patients that require extracorporeal cardiopulmonary resuscitation (eCPR) and its impact on severity of neuro-cognitive outcomes. A number of studies are focusing on development of predictive measures e.g., vitals data to identify patients at risk of cardiac, pulmonary, or hemodynamic events, heart rate variability for adverse events in the ICU, and pulse pressure variation is being monitored for post-op fluid responsiveness to identify patients needing fluid bolus.

The PIC iX clinical platform stores the most recent 38 weeks of the vitals data after which it is irretrievably purged. To retain data for research, the clinical system pushes data to STARR continuously, where it is processed to generate waveforms and vitals data in a non-proprietary format, and the resultant data can be linked to other data in STARR.

# Method

High density time-series bedside monitoring data are not well suited to traditional database and storage system technologies for long term storage and efficient retrieval of information. The raw data stored in a vendor proprietary system requires significant pre-processing prior to research use. In a recent paper [Goodwin2020], the authors propose a file structure for waveform data, a .tsc extension where the tsc file extension stands for time series compression. The authors also summarize previous work in the field of storage which rely on downsampling to a lower resolution data, and/or storing algorithmic derived variables. While this proposed approach retains high fidelity data using a domain specific compression, it is unwieldy for an academic medical center like Stanford to maintain yet another non-standard format. Our method creates the data repository in open source standards and relies on open source formats and accessible cloud technologies. The software application leading to creation of the STARR-Wave dataset has these main steps: extraction, database population, vitals/waves extraction, post processing of the extracted data and metadata cataloguing, as shown in Figure 1.

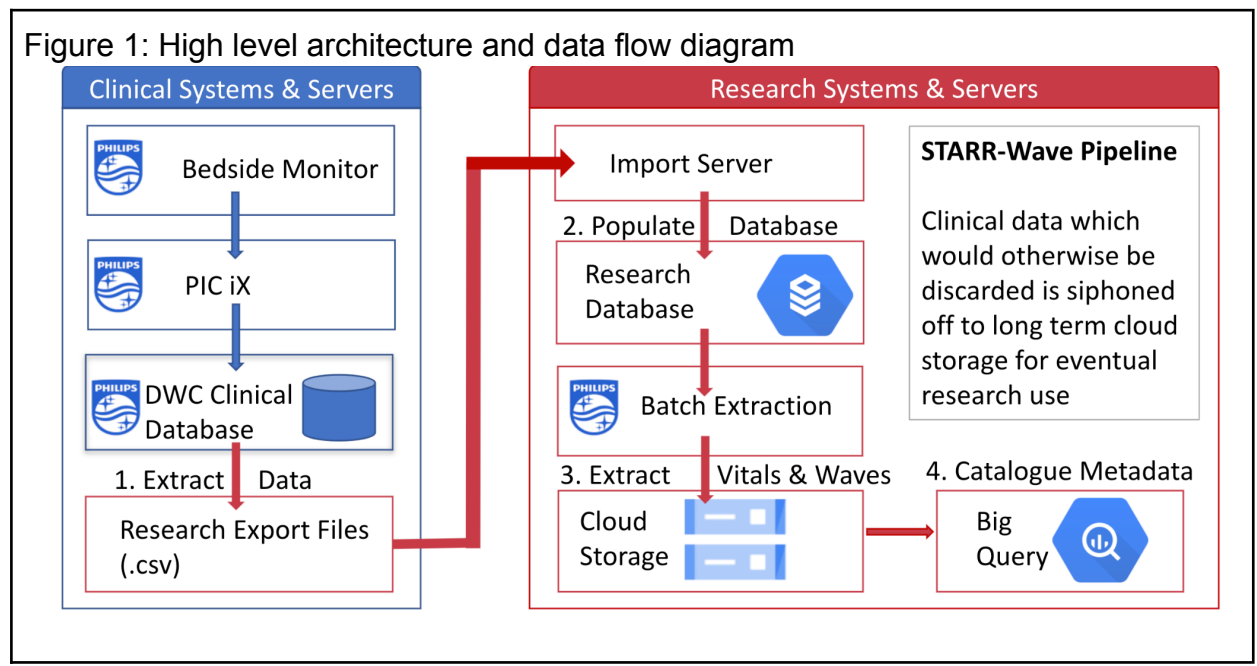

Figure 1: High level architecture and data flow diagram

Extraction is the only step that runs on the clinical system (Box 1 in Figure 1); it saves source data in chunks of 24 hours, midnight to midnight in a CSV format. The data content is proprietary to Philips and requires Philips software to interpret. The extracts are compressed and catalogued in a manifest containing file names, date time stamps, file sizes, and checksums. The nightly extract gets copied to a location in the cloud storage.

The remaining steps happen in our secure cloud. We use Google Cloud Platform and implement Stanford's minimum security guidelines for PHI data [Datta2020]. We use docker containers for software encapsulation. We also use on-demand Virtual Machines (VM) for data processing. When we started the project, we had three years of data to process, so we needed to boot up multiple VMs. But, in steady state, we do not need more than a single VM (see Table 3a). We use Google BigQuery (https://cloud.google.com/bigquery) for storing and retrieving metadata and cloud storage for storing and retrieving raw and waveform data.

The second step, database population (Box 2 in Figure 1), replicates the clinical DWC environment, downloads nightly extracts, decompresses them, checks their parity and reconstructs a SQL database schema (tables, views, indexes, procedures) according to the Philips DWC schema specifications. The Philips software used here is made accessible under a license sharing agreement between Stanford Children's Hospital and Stanford School of Medicine. Data is validated at this stage by verifying daily counts in the clinical database against the number of rows in the extracted files.

Step three is vitals/waves extraction (Box 3 in Figure 1). We run the Philips DWC toolkit batch extraction tool for extracting and pre-processing patient data studies in parallel. Each study, defined as upto 24 hours of data for a patient in a single bed, gets stored in a separate folder. This is the stage where we convert from the Philips DWC internal schema format to open standards format; the extraction format is PhysioNet (https://physionet.org/) compatible. The waveform data can be viewed by open-source viewers such as European Data Format (EDF) viewers [EDFViewers]. Some patients can be in a certain bed location for days, sometimes even weeks, while others change location frequently. Study length is limited to 24 hours maximum for a given patient and bed combination. Patient transfer triggers the creation of a separate study. This supports the use case where a researcher is only interested in vitals from certain beds or clinical units such as surgery or ICU.

The post processing step deals with compression of study folders, generation of study details, de-identification and ultimately the transfer of zipped studies to a cloud storage location.

Final step (Box 4, Figure 1) loads key data from the source system: metadata such as a study map containing bed location, study start/end times, study details (names and sizes of waves in that study); and a nightly waveform manifest containing compressed study folder names, sizes, and checksums.

Identified and de-identified vitals and metadata get stored in separate cloud storage locations/datasets. Some researchers' Institutional Review Boards (IRBs) might allow them to work with identified data whereas others may find it sufficient to work with de-identified data. Our solution allows us to offer self-service access to de-identified data; obtaining identified data remains a semi-automated process due to the requirement to verify compliance.

The study map metadata table documents if a given list of patients and/or bed labels and/or study times have any bedside monitoring data and records locations of cloud storage study folders. The study details metadata table can further aid researchers in only selecting studies that contain waveforms of specific interest.

The dataset can be linked to the institutional OMOP (Observational Medical Outcomes Partnership) data warehouse [Datta2020]. Users can start with a cohort in OMOP and then refine their cohort using study map metadata. Alternatively, they can start with a cohort using study map metadata and refine the cohort using the OMOP database.

The STARR-Wave application is containerized and can be scaled on demand depending on the load. Multiple instances can be started to process a large historical archive of extracts. Each instance processes one day worth of extracts and can operate independently.

## Results

In this section, we present data characterization as well as our software performance.

Here are the characteristics of the data in STARR-wave repository based on data from Feb 2017 to March 2021:
- Sampling frequency for PIC iX DWC captured data: 1 second vital numerics; up to 500 sample-per-second ECG waveforms.
- Total studies: ~620,000
- Compressed size of study folders: ~14 TB
- Average daily count of studies: ~400
- Total number of patients: ~48,000
- Average daily count of patients: ~280
- Uncompressed (Compressed) daily extract size: ~75 GB (~21 GB)
- Daily Philips database size: ~220 GB
- Average daily count of rows in alert table: ~180,000
- Average daily count of rows in wave sample table: ~10 million
- Average daily count of rows in enumeration value table: ~60 million
- Average daily count of rows in numeric value table: ~120 million

Figure 2 illustrates the rate of monthly accumulation of numeric vitals such as HR, $SPO_2$, BP from Feb 2017 until March 2021. During this time, the total number of monitors is ~600, however, the number of simultaneous monitors that export to DWC have varied between 350-450. This fluctuates as the monitors are turned off for a variety of operational reasons such as construction or activation of floating departments.The beds are retained due to the potential of being added back in the future as.

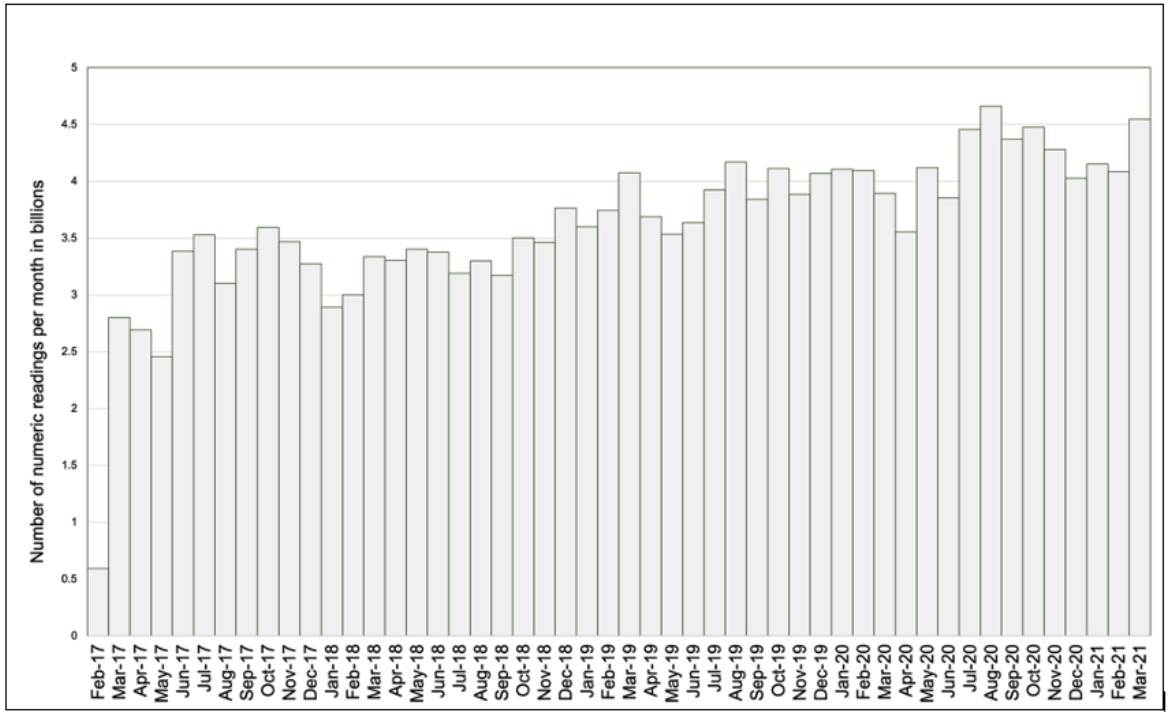

Figure 2: Numeric vitals in billions per month from Feb 2017 to March 2021

Table 1 provides a summary of waveform data from Feb 2017 until March 2021. Note that interpretation of the waveforms require deeper understanding of physiology in addition to device data documentation (https://med.stanford.edu/starr-wave/access.html#documentation). For example, ECG waveforms, abundant in our data, require understanding the location of various precordial leads, V1-V6, and how these measure different parts of the heart [Meek2002]. With a 5-electrode patient cable, there is only one possible V lead, and it can be placed anywhere; and thus labelled "V". Understanding of pediatric ECG is also important for age related changes in the anatomy and physiology of infants and children [Goodacre2002]. Table 2 illustrates the distribution of patients and studies in various clinical units by age groups.

| Wave Name | Symbol | Units of Measurement | Samples per second | Total Patients | Total Studies | Total Size (GB) |
|---|---|---|---|---|---|---|
| Airway Flow | AWF | l/min | 125 | 11,969 | 20,017 | 53 |
| Airway Oxygen | O2 | mmHg | 125 | 33,623 | 67,245 | 126 |
| Arterial Blood Pressure | ABP | mmHg | 125 | 7,139 | 47,557 | 604 |
| Arterial Blood Pressure | ART | mmHg | 125 | 3,029 | 16,369 | 230 |
| (Airway Expired) Carbon Dioxide | CO2 | mmHg | 125 | 34,961 | 120,085 | 799 |
| Central Venous Pressure | CVP | mmHg | 125 | 4,553 | 38,930 | 559 |
| Gas Analyzer Agent | AGT | % | 125 | 30,555 | 58,778 | 106 |
| Gas Analyzer Sevoflurane | SEV | % | 125 | 23,158 | 37,576 | 68 |
| Intra-Cranial Pressure | ICP | mmHg | 125 | 460 | 2,759 | 43 |
| Lead aVR - ECG Wave Label | aVR | mV | 500 | 10,850 | 47,746 | 1,101 |
| Lead I - ECG Wave Label | I | mV | 500 | 11,359 | 95,085 | 5,466 |
| Lead II - ECG Wave Label | II | mV | 500 | 47,222 | 495,329 | 17,750 |
| Lead III - ECG Wave Label | III | mV | 500 | 8,046 | 50,992 | 3,106 |
| Lead V - ECG Wave Label | V | mV | 500 | 11,068 | 54,259 | 1,411 |
| Chest/Percordial leads (V1-V6) | N/A | mV | 500 | 960 | 4,350 | 77 |
| Left Arterial Pressure | LAP | mmHg | 125 | 1,880 | 7,591 | 90 |
| Pleth Left Wave | PLETHl | N/A | 125 | 3,463 | 10,069 | 105 |
| Pleth Post Ductal | PLTHpo | N/A | 125 | 953 | 2,618 | 34 |
| Pleth Pre Ductal | PLTHpr | N/A | 125 | 1,694 | 4,961 | 57 |
| Pleth Right Wave | PLETHr | N/A | 125 | 6,103 | 15,140 | 142 |
| Pleth Wave | Pleth | N/A | 125 | 46,961 | 554,501 | 5,776 |
| Pleth wave from Telemetry | PlethT | N/A | 125 | 1,447 | 7,297 | 124 |
| Pulmonary Artery Pressure | PAP | mmHg | 125 | 1,880 | 5,137 | 51 |
| Resp Wave (Impedance via ECG electrodes) | Resp | Ohm | 63 | 40,058 | 490,755 | 2,729 |
| Right Arterial Pressure | RAP | mmHg | 63 | 1,062 | 8,505 | 130 |
| Umbilical Arterial Pressure | UAP | mmHg | 125 | 1,481 | 6,356 | 93 |
| Umbilical Venous Pressure | UVP | mmHg | 125 | 1,434 | 6,794 | 102 |

Table 1: Statistical summary of waveforms between Feb 2017 and March 2021. During this period, the total number of studies is 620,000 and total number of patients is ~48,000.

<sup>1</sup> Patients equal to or under the age of 28 days.
<sup>2</sup> Patients between the age of 29 days and less than 1 year.
* Some of the counts (such as those in ICN with age >=15 or Reproductive Infertility clinic age 1 to 4 years) are due to human errors in clinical systems.

| Clinical Unit | Neonates [1] | | Infants [2] | | 1 to 4 years | | 5 to 9 years | | 10 to 14 years | | 15 years or older | |
|---|---|---|---|---|---|---|---|---|---|---|---|---|
| | Patients | Studies | Patients | Studies | Patients | Studies | Patients | Studies | Patients | Studies | Patients | Studies |
| Anesthesiology | 364 | 466 | 1,039 | 1,661 | 6,344 | 12,947 | 5,683 | 11,104 | 4,755 | 9,272 | 9,986 | 15,398 |
| Cardiovascular ICU | 309 | 4,669 | 545 | 11,192 | 1,483 | 23,063 | 671 | 6,611 | 621 | 6,102 | 926 | 9,502 |
| Clinical Care Program | 21 | 58 | 30 | 72 | 137 | 386 | 70 | 265 | 329 | 4,703 | 743 | 10,778 |
| Intermediate Care Nursery | 2,181 | 23,748 | 607 | 12,802 | 120 | 2,855 | 0 | 0 | 0 | 0 | 4 | 4 |
| MRI/Radiology | 28 | 30 | 170 | 272 | 1,325 | 1,961 | 897 | 1,482 | 434 | 928 | 430 | 2,139 |
| Neonatal ICU | 3,332 | 31,890 | 876 | 19,636 | 233 | 5,797 | 0 | 0 | 1 | 41 | 6 | 7 |
| Surgery | 249 | 358 | 486 | 800 | 1,532 | 2,908 | 1,321 | 2,131 | 1,641 | 2,602 | 2,341 | 3,819 |
| Patient Care Unit | 425 | 1,831 | 1,446 | 12,042 | 5,652 | 48,312 | 3,581 | 30,288 | 3,741 | 34,013 | 4,205 | 47,847 |
| Pediatric ICU | 25 | 88 | 319 | 2,768 | 1,463 | 12,955 | 769 | 5,752 | 845 | 6,108 | 918 | 7,086 |
| Pre/Post Surgery | 32 | 46 | 654 | 1,009 | 5,407 | 10,833 | 5,165 | 11,659 | 4,616 | 11,370 | 4,679 | 12,218 |
| Reproductive Infertility Clinic | 0 | 0 | 0 | 0 | 1 | 1 | 0 | 0 | 0 | 0 | 2,277 | 6,933 |
| Stem Cell Transplant | 10 | 35 | 47 | 908 | 436 | 11,469 | 511 | 11,118 | 544 | 13,149 | 724 | 17,364 |

Table 2: Number of patients and studies by clinical unit and patient age group for data between Feb 2017 and March 2021.

This section describes the performance characteristics of running the STARR-Wave data processing pipeline. Each container runs on 8 core GCP VM instance with 150GB RAM, 300GB Standard Hard Disk Drive for operations, 300GB Solid State Drive for the database and local SSD (scratch space) for volatile data such as paging. The image consists of Windows Server 2019 Datacenter, SQL Server 2017 with Philips DWC SQL schema, Philips .NET application for extraction, GCP libraries and the STARR-Wave pipeline code developed at Stanford.

Table 3a presents sample pipeline execution cost including licensing fees for the VM host. At this time, we are using the on-demand windows server license option (https://cloud.google.com/compute/docs/instances/windows/ms-licensing-faq).

| Average daily Studies# | Average daily compressed Vital/Wave size | Time taken to process 24 hours data (~400 studies) | Processing Cost for 24 hours of data |
|---|---|---|---|
| 400 | 10GB | 4 hours | $8.5 |

Table 3a: Cost of a single day of data processing, Box 2-4 in Figure 1

Table 3b presents a yearly cost model for the STARR-Wave data lake. The estimated storage cost for raw yearly extracts and waveforms do not include egress costs for downloads or other cloud storage operations. Similarly, the estimated annual cost of storing the Big Data metadata does not include the cost of running queries.

| Average yearly raw extract size | Estimated annual cost for Cloud Object Storage for yearly raw extracts | Compute cost to process one year worth of data | Average yearly waveform size | Estimated annual cost of Cloud Object Storage for yearly waveforms | Estimated annual cost for Big Query metadata storage (50GB) |
|---|---|---|---|---|---|
| 7 TB | $1,716 | $3,103 | 3.5 TB | $864 | $12 |

Table 3b: Estimated total cost of processing an year of data and creating and storing the research usable assets. Total cost of storing one year's worth of raw proprietary data, processing that raw data into research usable format, storing the final waveforms and metadata in Cloud object storage and BigQuery is ~$5700.

One of the biggest challenges we encountered was that only half of the data in the source system has patient identifiers. To further complicate matters, vitals for one patient can be stored under multiple Philips internal patient ids. One example where this happens is when a patient moves from one bed to another and their internal id changes. A second confounding factor is that the same Philips internal patient id can be used to store data for multiple patients. An example of this is in surgery/OR rooms where patients come and go but the internal patient id for the vitals data stream remains the same. We used device logs containing patient encounter id, device/bed name, attach/detach timestamps and ADT (Admissions, Discharges and Transfer) from the EMR to line up patient movements, locations to the timestamps and location in the Philips DWC database to assign Medical Record Numbers (MRNs) to the extracted studies.

Philips databases come with their own set of challenges. Bed labels may not match up exactly with data in the EMR. For example, the bed named A13 in our EMR is labelled 13ALPHA in Philips PIC iX, and bed C01 is 01CHARLIE in Philips. A bed label mapping needed to be established and curated before it could be used. Timestamps are also not very well documented; patients are not admitted/discharged in Philips as often as one might expect, which makes it difficult to precisely determine a patient's stay in a given bed. In most cases we just have a patient id, bed label and a timestamp without a reliable admit/discharge status. In some cases, there will be more than one row for a patient id / bed label combination. These subsequent sets of rows were collapsed to get a time range for patient id and bed label. The time range from the EMR was checked for overlap with the time range in Philips for a matching bed to assign MRN. Lifetimeid (a.k.a. medical record number) entered in the Philips monitoring system was retained in the Study Map. Study start/end time for study generation were selected from EMR timestamps where available.

As a second pass, patient ids which have data that could not be matched previously are time-range compared to Admit Discharge Transfer (ADT) records in the EMR in order to assign an MRN. Patient ids that could not be assigned MRNs from both steps have their studies stored without an MRN.

When device logs are not available, we rely on ADT. However, ADT data has to be sanitized to remove noise before it is useful. Specifically, there are records in ADT that have the same admit/discharge timestamps for the same visit/encounter and bed. We ignore such rows.

| Event Id | Patient | Visit Id | Event | Bed | Event Time |
|---|---|---|---|---|---|
| 1 | John Doe | 1 | Transfer In | B09 | 2019-03-02 04:19:00 |
| 2 | John Doe | 1 | Transfer Out | B09 | 2019-03-02 04:19:00 |

There are duplicate records which need to be de-duplicated. .

| Event Id | Patient | Visit Id | Event | Bed | Event Time |
|---|---|---|---|---|---|
| 1 | John Doe | 1 | Admission | A17 | 2020-08-23 11:29:00 |
| 2 | John Doe | 1 | Admission | A17 | 2020-08-23 11:29:00 |
| 3 | John Doe | 1 | Transfer Out | A17 | 2020-09-23 20:24:00 |

There are records where patients are discharged and admitted right back to the same bed. We merge such rows by keeping the first admit and last discharge timestamp.

| Event Id | Patient | Visit Id | Event | Bed | Event Time |
|---|---|---|---|---|---|
| 1 | John Doe | 1 | Admission | A03 | 2020-07-02 15:59:00 |
| 2 | John Doe | 1 | Transfer Out | A03 | 2020-07-02 16:00:00 |
| 3 | John Doe | 1 | Transfer In | A03 | 2020-07-02 16:00:00 |
| 4 | John Doe | 1 | Transfer Out | A03 | 2020-07-02 16:16:00 |
| 5 | John Doe | 1 | Transfer In | A03 | 2020-07-02 16:16:00 |
| 6 | John Doe | 1 | Discharge | A03 | 2020-07-02 21:03:00 |

The goal of the sanitizing process is to end up with a single ADT record with patient visit/encounter id, bed name, and admit/discharge timestamps, similar to a device log entry. This approach enabled us to assign MRNs to 75% of patient ids missing LifeTimeId in the source system with an accuracy of 92% or more. Having patient identifiers on the vitals data significantly improves its utility in medical/translational research as the MRN can be used to join with the EMR and with data from other ancillary medical systems.

# Acknowledgments

The STARR suite is made possible by the Stanford School of Medicine Dean's Office. Using CRediT taxonomy (https://casrai.org/credit/), we present the contributing roles for our authors – Sanjay Malunjkar (Conceptualization, Methodology, Data curation, Formal Analysis, Investigation, Software, Validation), Susan Weber (Supervision, Resources, Investigation, Editing), Somalee Datta (Writing - review & editing), Priya Desai (Product Management), and Joe McCullagh (Project Management). We would like to thank Eric Helfenbein, Principal Scientist, Philips Healthcare for vendor support and close partnership, Sijo Thomas, Clinical Program Manager & Carlos DeSousa, Systems Engineer, Stanford Children's Health for IT liaison. We acknowledge support of members of Research IT team (https://med.stanford.edu/researchit), Dr. Todd Ferris (Project Ideation), Chief Technology Officer, Technology & Digital Solutions and Michael Halaas (Funding acquisition), Deputy Chief Information Officer of Technology & Digital Solutions and Associate Dean of Industry Relations at Stanford School of Medicine.